\documentclass[fleqn,10pt]{wlscirep}
\usepackage[utf8]{inputenc}
\usepackage[T1]{fontenc}
\title{Network analysis of a complex disease: the gut microbiota in the inflammatory bowel disease case}

\author[1]{Mirko Hu}
\author[2,*]{Guido Caldarelli}
\author[3]{Tommaso Gili}
\affil[1]{University of Parma, Department of Medicine and Surgery, Parma, 43121, Italy}
\affil[2]{Ca’ Foscari University of Venice, Department of Molecular Science and Nanosystems, Venice, 30123, Italy}
\affil[3]{IMT School for Advanced Studies Lucca, Networks unit, Lucca, 55100, Italy}

\affil[*]{guido.caldarelli@unive.it}

\begin{abstract}
Inflammatory bowel diseases (IBD) are complex diseases in which the gut microbiota is attacked by the immune system of genetically predisposed subjects when they are exposed to yet unclear environmental factors. The complexity of this class of diseases makes them suitable to be represented and studied with network science. In the project, the metagenomic data of the gut microbiota of control, Crohn’s disease, and ulcerative colitis subjects were divided in three ranges (prevalent, common, uncommon). Then, correlation networks and co-expression networks were used to represent this data. The former networks involved the calculation of the Pearson’s correlation and the use of the percolation threshold to binarize the adjacency matrix, whereas the latter involved the construction of the bipartite networks and the monopartite projection after binarization of the biadjacency matrix. Then, centrality measures and community detection were used on the so-built networks. The main results obtained were about the modules of \textit{Bacteroides}, which were connected in control subjects’ correlation network, \textit{Faecalibacterium prausnitzii}, where co-enzyme A became central in IBD correlation networks and \textit{Escherichia coli}, which module has different position in the different diagnoses networks. 
\end{abstract}
\begin{document}

\flushbottom
\maketitle
\thispagestyle{empty}

\noindent % Please note: Abbreviations should be introduced at the first mention in the main text – no abbreviations lists. Suggested structure of main text (not enforced) is provided below.

\section*{Introduction}

% The Introduction section, of referenced text\cite{Figueredo:2009dg} expands on the background of the work (some overlap with the Abstract is acceptable). The introduction should not include subheadings.

Microbes are ubiquitous. From radioactive waste to the human gastrointestinal tract, they can be found nearly everywhere. In and on the human body, they have evolved to co-exist with their host and it is estimated that the amount of microbes hosted by the human body is of the same order of magnitude as the number of human cells \cite{Sender2016RevisedBody}. In particular, the $10^{14}$ commensal microbes living in the intestinal tract form the human gut microbiota, which has evolved to live in symbiosis with its host \cite{Thursby2017IntroductionMicrobiota}. It is widely accepted that this symbiosis begins from birth and the microbial communities stabilize with age until the formation of an adult microbiota \cite{Tanaka2017DevelopmentLife}. Its genetic content (called the microbiome) characterizes each individual, rising also concerns about identity and privacy issues, specifically when the study and the manipulation of the microbiota are considered \cite{Ma2018HelpApplications}. Since the 1840s, when the concept of gut microbiota first appeared, the topic has been studied for two centuries \cite{Farre-Maduell2019TheNissle}, and, at the moment, it is known that the gut microbiota has a fundamental role in shaping the gut barriers \cite{Natividad2013ModulationImplications}, training the host immune system and regulating the metabolism \cite{Sommer2013ThePhysiology}. When the compositional and metabolic equilibrium of the commensal microbes living in the gut is disrupted, different types of diseases arise such as metabolic disorders or central nervous system disorders \cite{Belizario2018MicrobiomeDysbiosis}. Historically, traditional medicine attempted to re-establish this equilibrium through remedies intervening on the digestive system, like fasting, diets, assumption of spring waters or laxatives. A quite recent procedure introduced to tackle the \textit{C. difficile} infection is the faecal microbiota transplantation (FMT) \cite{Borody2012FecalApplications} which consists in repopulating the intestinal tract of an ill subject with the microbiota of a healthy donor. 

Inflammatory bowel diseases (IBDs), which comprise Crohn's disease (CD) and ulcerative colitis (UC), are an important class of diseases that arises from dysbiosis and are being treated with FMT. Typical symptoms of this class of diseases are chronic diarrhoea, abdominal pain, rectal bleeding, weight loss and fatigue \cite{Singh2011CommonDiseases}. Although CD and UC are both characterized by the inflammation of the intestinal tract, there are several differences between the two diagnoses that span from the environmental factors that cause them, e.g. smoking or diet, to the clinical and endoscopic findings in the two diagnoses \cite{Ananthakrishnan2017InflammatoryGuide}. Overall, IBDs are becoming widespread in modern society because of the change in lifestyle, socioeconomic developments and environmental causes \cite{Manichanh2012TheIBDc}. Until now, it is known that IBD is an exaggerated immune response to the gut microbiota of genetically predisposed subjects under the influence of the external environment. This complex interplay between genetics, the microbiota, the immune system and the environment makes it particularly hard to understand this class of diseases.
\cite{Kirsner1988HistoricalDisease.} offered a complete historical review of the IBD until the 1980s, by quoting Hippocrates who described diarrhoea as a symptom of an infectious (or non-infectious) disease to a description of hypothetical pathogenesis of IBD, which the microbiota was not considered, though. A more recent projection predicted the evolution of the disease between 2015 and 2025 and updated the possible origins of IBD including the action of antibiotics on the gut microbiota in Western society \cite{Kaplan2015The2025}. \cite{Xavier2007UnravellingDisease} summarized the findings of the origins of the IBD, mentioning the complexity of the disease. Another historical review focuses on the genetics of the IBD \cite{Lees2009GeneticsHistory} identified \textit{NOD2} as the first CD susceptible gene and then described the evolution of the IBD genetics with the coming of the modern genome-wide association study. One of the first most comprehensive work describing the interaction of all the aforementioned factors can be found in \cite{Zhang2014InflammatoryPathogenesis}. Whereas, the systems biology approach to the study of IBD was presented by \cite{Fiocchi2021IBDStay}, which proposed the creation of an IBD interactome, a complex system connecting all the potential agents interacting among them that derived from the combination of different omics \cite{DeSouza2017TheTherapy}.

Our work starts from here and attempts to provide tools and methods from network science useful to build and to study the IBD interactome with a systems biology approach by commencing from the metagenomic data of the gut microbiome. This approach is typical of network medicine, a novel discipline that mixes network science with systems biology to tackle the challenges offered by the progress of personalized medicine \cite{Barabasi2011NetworkDisease}, which opposes to the current effective yet drastic procedures like the aforementioned FMT. Network science is the discipline used to analyse complex systems and could be suited to understand a complex disease like IBD in which a complex system like the gut microbiota plays a fundamental role. Complexity in the intestinal microbial communities arises at different scales; from the macroscopic point of view, we have the ecological interactions \cite{Faust2018SignaturesSeries, Bucci2016MDSINE:Analyses} that describe the relationships among the species in the gut microbiota; among these, we have three main different types of interactions \cite{Coyte2019UnderstandingMicrobiome}; positive interactions (cooperation, commensalism, cross-feeding), negative interactions (competition, ammensalism), and asymmetric interactions (exploitation, predation, parasitism). Going towards a more microscopic scale, we can find the gene networks, often represented by gene co-expression networks \cite{Vernocchi2020NetworkCancer} and metabolic networks built by connecting the substances, known as metabolites, reacting in the same metabolic processes \cite{Bauer2018FromMicrobiota}. When the interaction between bacterial proteins is considered, we are dealing with metaproteomics, which is a novel tool in the analysis of IBD; nonetheless, the data used is still scarce \cite{Segal2019TheDisease}.

The application of network science for the study of the complexity of the gut microbiome is recent and one of the first research was in the case of \textit{C. difficile} infection \cite{Stein2013EcologicalMicrobiota}. The microbiome in this work was represented as a boolean network derived from binarized temporal data of the abundance of specific bacteria species in the gut. Although the study was able to capture the dynamics of the bacterial species, e.g. negative, positive or neutral interaction, it did not take into account the genetic expression of the microbiome (metagenome), which could explain better the complex interplay between the bacterial species. Our study, by contrast, gives a static screenshot of the microbial interactions through metagenomics. A more recent study \cite{Chen2020GutObesity} analysed the co-abundance network built with SparCC \cite{Friedman2012InferringData}, the need of this tool is due to the necessity of sparsifying the network that would have too many correlated nodes because of normalization and a $p$-value threshold too high \cite{Weiss2016CorrelationPrecision}. Based on a topological property of the biological networks, the work by \cite{Vernocchi2020NetworkCancer} portraits a weighted gene co-expression network analysis by building a network from metagenomic data and removing the weaker edges based on the assumption that the final network would be scale-free. In our work, we used thresholding methods that rely on the network topology such as the percolation threshold or the $p$-value for the projected edges, similarly to the later research. These methods should overcome the aforementioned problems.

\section*{Results}

% Up to three levels of \textbf{subheading} are permitted. Subheadings should not be numbered.

In the next paragraphs, the pathways are called with their node numbers for the sake of brevity and clarity, it is possible to consult the table mapping these correspondences in the Supplementary Material.

\subsection*{Bacteroides and Faecalibacterium prausnitzii}

In the prevalent pathways, the number of edges in the NI correlation network was $3356$ ($\text{th}_\text{NI} = 0.453$), in CD correlation network was $2905$ ($\text{th}_\text{CD} = 0.357$), and in UC correlation network was $3160$ ($\text{th}_\text{UC} = 0.364$). The results showed that NI metagenome was more connected for prevalent pathways and the percolation threshold was higher compared to the percolation thresholds in CD and UC correlation networks, translating into more strongly correlated nodes in the NI correlation network. From the community detection algorithm, we obtained that the lowest modularity ($0.538$) can be found in NI correlation network, meaning that it was not possible to completely separate some of the modules and there would be interconnections between them, CD and UC correlation networks resulted in a modularity of $0.583$ and $0.622$, respectively.

As we can see in Figure \ref{res1}, the pathways in the NI correlation network were divided into three large modules, one module was isolated, by contrast, the other two modules were communicating strictly through several nodes. The isolated module was composed of \textit{Bacteroides vulgatus} and \textit{Bacteroides uniformis} pathways, this meant that in control subjects the two species co-variated and were interdependent through specific pathways (on the frontiers of the species modules, it was possible to find nodes $794$, $658$, $292$, $477$ on the \textit{B. uniformis} side and nodes $261$, $855$, $1027$, $1068$ on the \textit{B. vulgatus} side). The light purple module, on the other hand, contained \textit{Faecalibacterium prausnitzii}, whereas the remaining large module contained the \textit{B. ovatus} pathways, \textit{E. rectale} pathways, and the unclassified species pathways. The nodes with the highest betweenness centrality among the unclassified pathways in the two connected modules were connected through: 
\begin{enumerate}
    \item node $821$ (PWY-6527: stachyose degradation);
    \item node $177$ (GLYCOGENSYNTH-PWY: glycogen biosynthesis I (from ADP-D-Glucose));
    \item node $1225$ (PWY66-422: D-galactose degradation V (Leloir
pathway));
    \item node $751$ (PWY-6317: galactose degradation I (Leloir
pathway)).
\end{enumerate}
Whereas, the nodes with the highest betweenness centrality among the \textit{F. prausnitzii} pathway module that was connected to the unclassified pathway module were:
\begin{enumerate}
    \item node $466$ (PWY-5659: GDP-mannose biosynthesis);
    \item node $724$ (PWY-6277: superpathway of 5-aminoimidazole ribonucleotide biosynthesis);
    \item node $579$ (PWY-6121: 5-aminoimidazole ribonucleotide biosynthesis I);
    \item node $610$ (PWY-6122: 5-aminoimidazole ribonucleotide biosynthesis II).
\end{enumerate}
To notice that pathway of node $466$ correlated with only  $4$ \textit{E. rectale} pathways.

In the CD correlation network, there were fewer connections and the network was divided into $6$ modules. Each module corresponded to the species groups of pathways. The smallest module was composed of \textit{E. rectale} pathways. The largest (light purple) module comprising \textit{F. prausnitzii} was connected to the unclassified (green) module by means of node $105$ similarly to the UC correlation network. Moreover, an additional bridge connecting node was node $1261$, which was linked to nodes $821$ and $204$, to mention two high betweenness centrality nodes among the unclassified pathways. High betweenness centrality nodes $137$ and $462$ connected unclassified species module with \textit{B. ovatus} module, node $632$, in turn, was connected to node $988$ linking \textit{B. ovatus} module to the \textit{B. uniformis} module. Finally, similarly to the NI correlation network, \textit{B. vulgatus} and \textit{B. uniformis} were connected though (nodes $370$, $1042$, and $1074$ on the former side and nodes $1205$ and $5$ on the latter side).

The pathways in the UC correlation network were divided into $5$ modules. The smallest module (coral red) was composed of \textit{E. rectale} pathways scattered around the network. The dark green nodes mixed to the turquoise nodes were \textit{B. ovatus} pathways mixed with \textit{B. uniformis} pathways, respectively. The green module comprised of unclassified species pathways, whereas, the light purple module comprised of \textit{F. prausnitzii} pathways. The green and the light purple module were strictly connected similarly to the NI correlation network, the pathways connecting them were the node $105$, which was linked to several nodes of both modules and node $579$, which was linked to nodes $1093$ and $1284$. Even \textit{E. rectale} behaved as a bridge between the two aforementioned large modules through a few connections. Furthermore, there was one node of \textit{F. prausnitzii} module that was deeply correlated with all the \textit{B. vulgatus} pathways: node $133$. Finally, two nodes with high betweenness centrality were node $901$ in the light purple module and node $873$ in the mixed dark green and turquoise module. Differently from the NI correlation network, \textit{B. uniformis} did not correlate \textit{B. vulgatus}, whereas it correlated with \textit{B. ovatus}.

\begin{figure}[ht]
\centering
\includegraphics[width=10.5 cm]{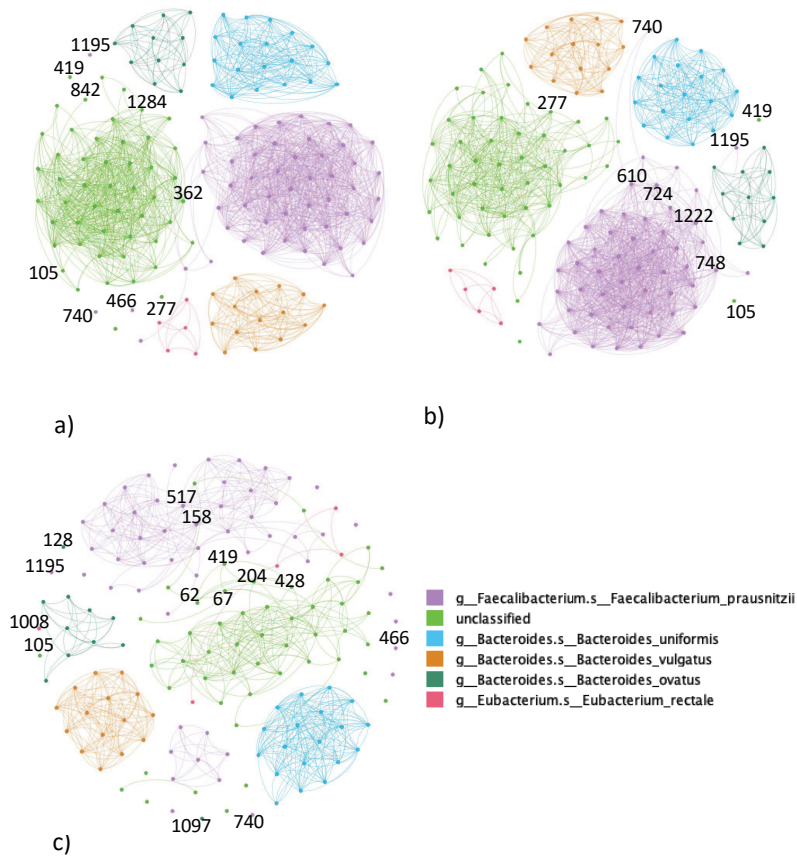}
\caption{Projected network representation of prevalent metagenomic pathways in (\textbf{a}) NI subjects, (\textbf{b}) CD subjects, (\textbf{c}) UC subjects. Networks are obtained through a bipartite projection and the exclusion of an edge between two nodes is made through a comparison with a null model. The nodes highlighted describes the pathway with highest betweenness centrality. The legend remarks the species of each node/pathway.\label{res2}}
\end{figure}

In Figure \ref{res2}, we have projected the bipartite network onto the nodes of the bacterial pathways and we validate the projection through a null model ($\alpha=0.05$ and $fwer = none$). Again, we divided the pathways according to their presence along with the samples. We considered the case of 75\% of the presence across the samples for NI subjects. We found $1715$ edges for $153$ nodes and the community detection resulted in $6$ large communities; namely, the unclassified species community, the \textit{F. prausnitzii} community, the \textit{E. rectale} community, the \textit{B. uniformis} community, the \textit{B. ovatus} community and \textit{B. vulgatus} community. All the communities identified were isolated, the node $105$ was connected to the unclassified module through the nodes $842$, $362$, and $1284$. Also, nodes $419$ and $277$ were separated from the rest of the unclassified module. Considering \textit{F. prausnitzii} module, nodes $1195$, $740$, and $466$ were disconnected from the rest of the module.

In the CD case, there were $153$ nodes and $1615$ edges, the community detection algorithm identified $6$ different modules, one for each bacterial species identified in the previous cases. All the communities were isolated without nodes connecting them. Differently from the NI case, node $105$ does not result connected to the rest of the unclassified module. Similarly to the NI case, also node $419$ was not isolated, whereas node $277$ was well connected to the rest of the unclassified module. When \textit{F. prausnitzii} was considered, we obtained that node $1195$ was connected to nodes $748$ and $1222$, that both were well connected to the \textit{F. prausnitzii} pathways. Also, node $740$ differently from the NI projected network is connected to two nodes of the bacterium module, namely $610$ and $724$, which were both pathways involving  5-aminoimidazole ribonucleotide. On the other hand, node $466$ was connected to node $538$ and to nodes, $748$ and $1222$, where the former was involved in the biosynthesis of urate and the latter were involved in the degradation of galactose.

In the UC case, we could find $153$ nodes and $669$ edges. Differently from the previous cases, the community detection algorithm could not isolate modules corresponding to the $6$ species of the prevalent pathway group. The \textit{F. prausnitzii} module was split into two modules held together by the nodes $158$ and $517$. Similarly to the CD projected network, node $105$ was isolated from the rest of the unclassified module, whereas node $419$ was connected to the rest of the module by means of nodes $62$, $204$, $428$, and $67$. Differently from the NI projected network and similarly to the CD projected network, in the UC projected network, the node $277$ was well connected to the rest of the unclassified module. When the \textit{F. prausnitzii} module was considered, nodes $1195$, $740$ and $466$ were isolated from the rest of the module in the same fashion as the NI case. Furthermore, the UC projected network by displaying fewer connections had more isolated pathways compared to the NI and the CD case. For instance, nodes $128$ and $1097$ detached from the \textit{B. ovatus} module, node $1008$ detached from the \textit{E. rectale} module, and several more fragments of the module.

\begin{figure}[ht]
\centering
\includegraphics[width=10.5 cm]{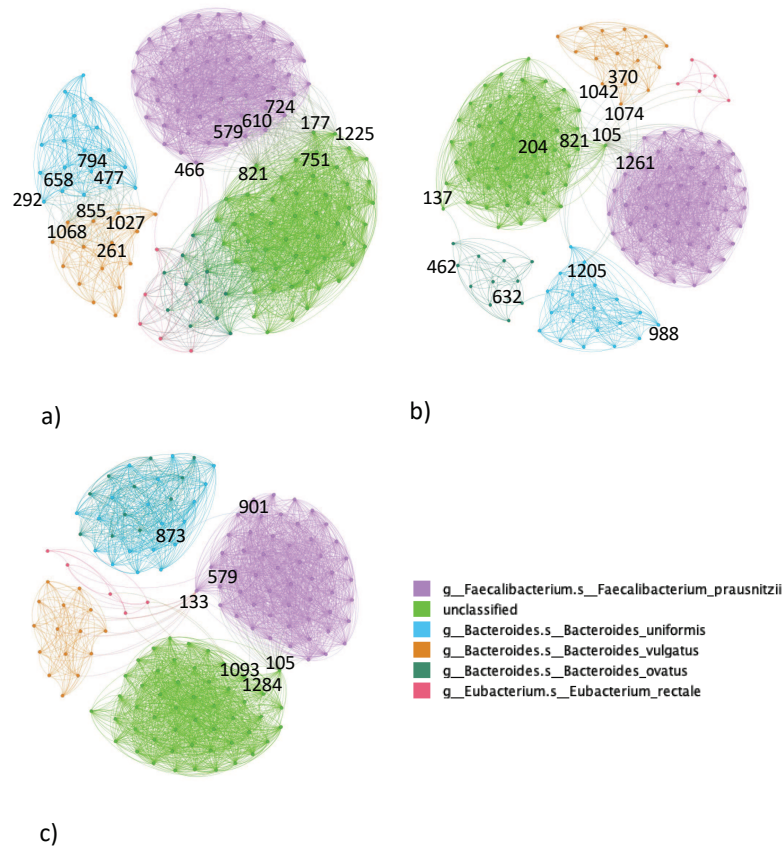}
% \caption{Legend (350 words max). Example legend text.}
\caption{Correlation networks of prevalent metagenomic pathways in (\textbf{a}) NI subjects, (\textbf{b}) CD subjects, (\textbf{c}) UC subjects. In Figure (\textbf{a}), nodes in the frontier between \textit{B. vulgatus} and \textit{B. uniformis} and in the frontier between \textit{F. prauznitzii} and unclassified module are highlighted. In Figure (\textbf{b}) and Figure (\textbf{c}), nodes with high betweenness centrality like node 105 are higlighted. The legend remarks the species of each node/pathway.}
\label{fig:res1}
\end{figure}

\subsection*{Escherichia coli}

\begin{figure}[ht]
\centering
\includegraphics[width=10.5 cm]{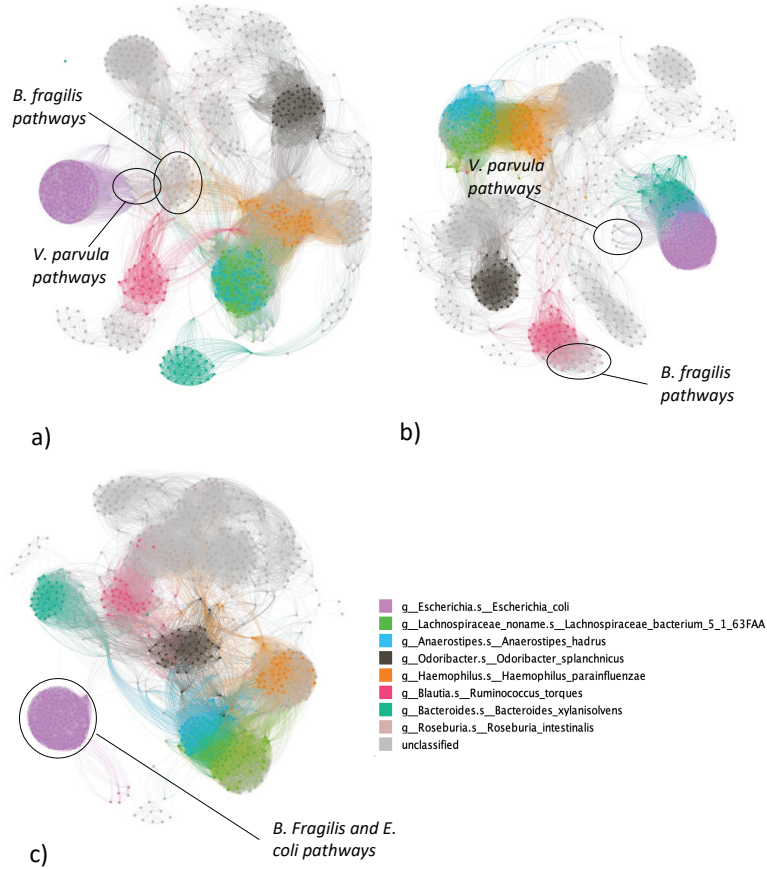}
\caption{Correlation networks of uncommon metagenomic pathways in (\textbf{a}) NI subjects, (\textbf{b}) CD subjects, (\textbf{c}) UC subjects. The nodes encircled shows the position of \textit{B. fragilis} and other species like \textit{E. coli} and \textit{V. parvula}. The legend remarks the species of each node/pathway, where the grey nodes are classified as 'other'.\label{res3}}
\end{figure}

As it is possible to observe in Figure \ref{res3}, the relatively uncommon pathways were $910$ in total. There were $49203$ edges in NI correlation network ($\text{th}_\text{NI} = 0.474$),  $42617$ edges in CD correlation network ($\text{th}_\text{CD} = 0.446$), $44523$ edges in UC correlation network ($\text{th}_\text{UC} = 0.465$). The number of edges was comparable in the three networks. We could observe nine modules in the NI correlation network (modularity $0.623$, nine modules in the CD correlation network (modularity $0.644$), and six modules in the UC correlation network (modularity $0.695$). In every network, it was possible to identify an approximately isolated ball-shaped module containing \textit{E. coli} pathways. It was interesting to highlight the position of \textit{B. fragilis} pathways in respect of \textit{E. coli} pathways in the different diagnoses. In the NI correlation network, \textit{B. fragilis} pathways were connected to the \textit{E. coli} module through \textit{V. parvula} pathways, by contrast, in the UC correlation network, \textit{B. fragilis} pathways were incorporated and surrounded by the same module containing \textit{E. coli} pathways, whereas in the CD correlation network the bacterial pathways of the two species were completely separated. In the NI correlation network, the module containing \textit{E. coli}  included also bacterial pathways of other species, notably \textit{E. siraeum} and \textit{R. gnavus} pathways that behaved as bridge nodes between \textit{E. coli} containing module and the rest of the network. In the CD correlation network, this role was assumed by \textit{R. intestinalis} and \textit{V. parvula}, whereas, in the UC correlation network, we did not observe any pathways behaving as bridges nodes. An additional important module that was also the largest was the one mainly composed of \textit{R. torques}, \textit{A. hadrus}, \textit{L. bacterium 5 1 63FAA}, plus other minor species. The pathways belonging to the last two mentioned species were strictly intertwined forming the second largest ball-shaped group of nodes. The same ball-shaped group of nodes was present in the CD correlation network, by contrast, in the UC correlation network, in the UC correlation network the two species were in the same module and the nodes were not mixed in the ball but they laid separately. 

\begin{figure}[ht]
\centering
\includegraphics[width=10.5 cm]{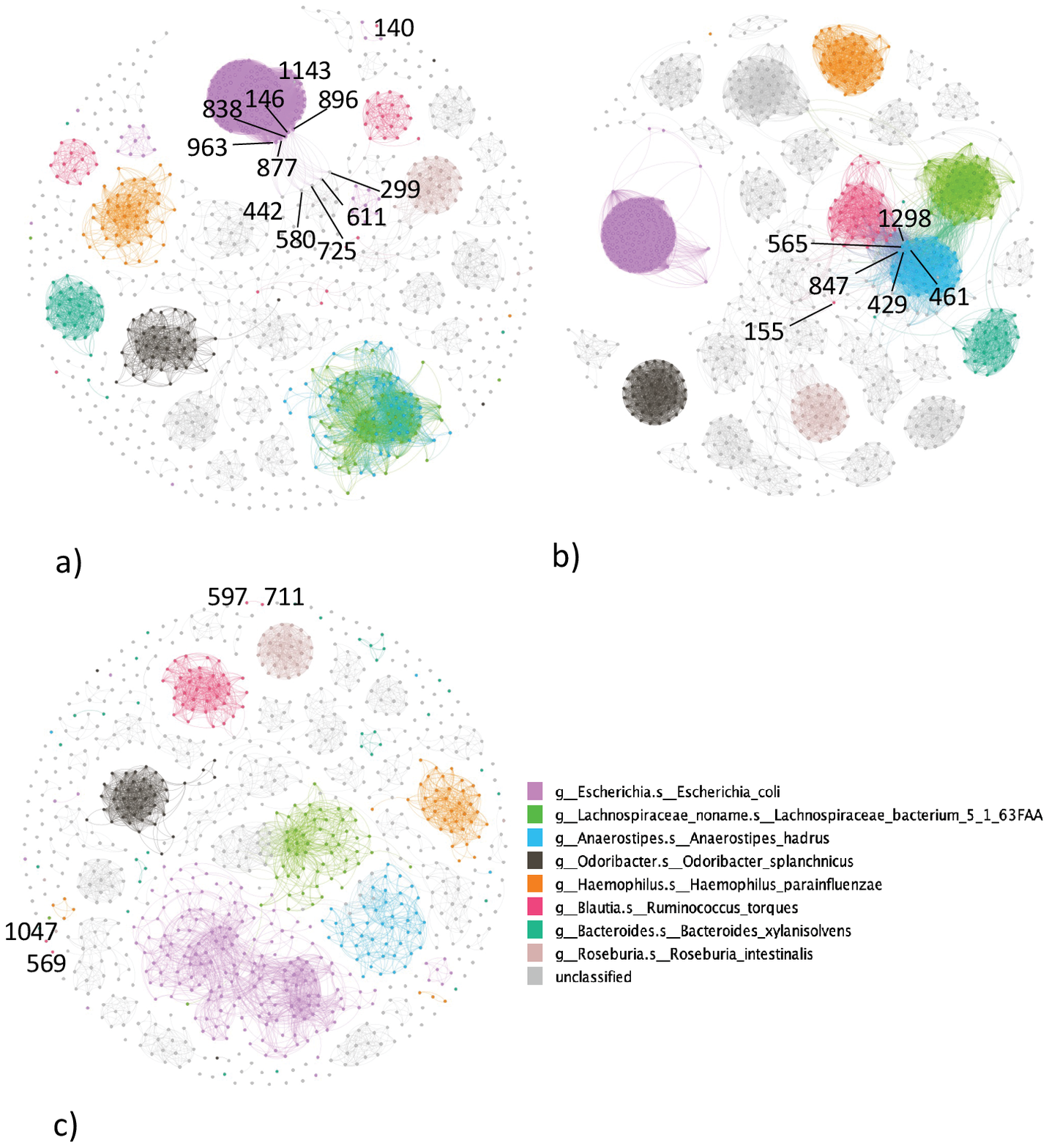}
\caption{Projected network representation of uncommon metagenomic pathways in (\textbf{a}) NI subjects, (\textbf{b}) CD subjects, (\textbf{c}) UC subjects. Networks are obtained through a bipartite projection and the exclusion of an edge between two nodes is made through a comparison with a null model. The nodes highlighted corresponds to nodes with high betweenness centralities or to nodes that compare to the NI projected network are isolated. The legend remarks the species of each node/pathway.y, where the grey nodes are classified as 'other'.\label{res4}}
\end{figure}

In the NI network projection, we obtained a network with $953$ nodes and $15394$ edges. As it is possible to notice in Figure \ref{res4}, there were several modules that emerged from the network. The most evident was the \textit{E. coli} module which was one of the modules that composed the \textit{E. coli} group in the uncommon pathways. This bigger module was connected to \textit{F. plautii} via $5$ nodes on the \textit{E. coli} side and via $5$ nodes on the \textit{F. plautii} side. On the \textit{E. coli} side:
\begin{enumerate}
    \item node $146$ (FUCCAT-PWY: fucose degradation);
    \item node $896$ (PWY-6737: starch degradation V);
    \item node $838$ (PWY-6609: adenine and adenosine salvage III);
    \item node $877$ (PWY-6703: preQ0 biosynthesis);
    \item node $963$ (PWY-7199: pyrimidine deoxyribonucleosides salvage).
\end{enumerate}
On the \textit{F. plautii} side: 
\begin{enumerate}
    \item node $442$ (PWY-5188: tetrapyrrole biosynthesis I (from glutamate));
    \item node $611$ (PWY-6122: 5-aminoimidazole ribonucleotide biosynthesis II);
    \item node $725$ (PWY-6277: superpathway of 5-aminoimidazole ribonucleotide biosynthesis);
    \item node $299$ (PEPTIDOGLYCANSYN-PWY: peptidoglycan biosynthesis (meso-diaminopimelate containing));
    \item node $580$ (PWY-6121: 5-aminoimidazole ribonucleotide biosynthesis I).
\end{enumerate}
Curiously there was also a \textit{C. bolteae} pathway well-connected to the \textit{E. coli} main module (node $1143$). The remaining part of the \textit{E. coli} group was represented by fatty acid metabolism pathways (node $140$) or pathways involving mannose biosynthesis, were connected to other species pathways. Apart from these nodes scattered around the network, the species group had two main ball-shaped modules thanks to the well-connected nature of the nodes inside the module. Another interesting community was the module formed by the nodes of two different species \textit{Lachnospiraceae bacterium 5 1 63FAA} and \textit{A. hadrus}. The edges connecting the nodes of the two species were so dense that the two groups formed a unique module.

In the CD network projection, there were $953$ nodes and $24298$ edges. One of the immediately visible properties of the projected network was the isolated module composed of \textit{E. coli} pathways. Compared to the NI case, there were no connections to the other species nodes. \textit{R. torques} module was connected to \textit{A. hadrus} module, which in turn was connected to \textit{L. bacterium 5 1 63FAA} module. $5$ nodes of \textit{A. hadrus} community were connected to most of the nodes of the other two species exhibiting high betweenness centrality, they were:
\begin{enumerate}
    \item node $1298$ (VALSYN-PWY: L-valine biosynthesis);
    \item node $429$ (PWY-5104: L-isoleucine biosynthesis IV);
    \item node $461$ (PWY-5659: GDP-mannose biosynthesis);
    \item node $847$ (PWY-6700: queuosine biosynthesis);
    \item node $565$ (PWY-6121: 5-aminoimidazole ribonucleotide biosynthesis I).
    \end{enumerate}
Similarly to the NI case, node $155$ was connected only to pathways of other species. Another difference with the NI projected network was the edges of \textit{R. intestinalis} which in this case were connected to \textit{B. instestinihominis}, whereas, in the NI case, they were connected to nodes $671$ and $41$.

The UC network projection had $953$ and only $5432$ edges. The first striking property of this network was that there were much fewer edges compared to the previous two cases. It was not possible to say much about the module of \textit{E. coli}. On the other hand, it was possible to observe that \textit{L. bacterium 5 1 63FAA} and \textit{A. hadrus} were completely separated. Furthermore, \textit{L. bacterium 5 1 63FAA} module was connected to the \textit{R. hominis} module through several nodes. Related to \textit{R. torques} group, we had one main module similar to the other cases, two isolated nodes (node $1047$ as the NI case and node $569$) and two nodes connected between them (nodes $711$ and $597$), but separated from the rest of the pathways. 

\unskip

% \subsection*{Subsection}

% Example text under a subsection. Bulleted lists may be used where appropriate, e.g.

%\begin{itemize}
%\item First item
%\item Second item
%\end{itemize}

%\subsubsection*{Third-level section}
 
%Topical subheadings are allowed.

\section*{Discussion}

In the literature, the reduction of \textit{F. prausnitzii} has been associated with IBD \cite{Cao2014AssociationLiterature}, nevertheless, in our results we recognized that instead of a decrease in the quantity of \textit{F. prausnitzii} pathways expressed, there was a change of the wiring in the metagenomic network. In particular, what changed from the NI correlation network and the IBD correlation networks were the bridge pathways connecting the module of unclassified pathways and the module containing the \textit{F. prausnitzii} pathways. For instance, in the NI correlation network the bridge pathway was node $821$, whereas in the IBD networks the bridge pathways between the two modules were node $105$. The correlation between these bridge nodes and the nodes in the aforementioned modules meant that both modules relied on the pathway to function correctly. On one hand, the central substance was the tetrasaccharide stachyose which in the pathway is degraded into UDP-alpha-D-glucose and has been recognized as potential probiotic against enterotoxigenic \textit{E. coli} \cite{Xi2020EffectsColi}; on the other hand, there was the coenzyme A, which has a fundamental role in the metabolism and, in particular, it is important in the oxidation of fatty acids. Several studies linked the alteration of fatty acid production to the IBDs, hence, this change in the centrality of the pathway related to this substance could be investigated further to explain the origins of IBDs \cite{Xi2020EffectsColi}. The fact that the modules of \textit{Bacteroides} in IBD networks corresponded to its species whilst in NI network are gathered in one module could demonstrate that in IBD the different \textit{Bacteroides} species proliferates the gut independently. This could confirm the meta-analysis by \cite{Zhou2016LowerMeta-analysis}, who showed that lower levels of \textit{Bacteroides} were associated with IBDs. Other pathways differentially wired between NI and IBD networks are those involving the bacterial metabolite 5-aminoimidazole ribonucleotide (nodes $579$, $610$, $724$), these nodes were behaving as bridges in the NI correlation network, by contrast, in the IBD correlation networks, they were substituted by a unique bridge node (node $105$).

In the range of uncommon bacterial species, we can observe the \textit{E. coli} module; in the literature, this bacterial species has a recognized role in the development of IBD \cite{Rhodes2007TheDisease, Schirmer2019MicrobialDisease}. It was possible to observe the different interplay between \textit{E. coli}, \textit{V. parvula} and \textit{B. fragilis} across the different diagnoses. The increase of \textit{E. coli} and \textit{B. fragilis} in IBD was observed in a previous study \cite{Keighley1978InfluenceMicroflora.}, but our results provide an additional information about the differential wiring scheme of the aforementioned species. In particular, it seemed that \textit{V. parvula} pathways mediated the connection of \textit{E. coli} with the other module in the correlation network. In particular, in NI correlation network, \textit{V. parvula} pathways were in the same module of \textit{B. fragilis} pathways which were connected to the rest of the correlation network. In the CD correlation network, \textit{V. parvula} pathways were included in the \textit{E. coli} module, just to remark how close the two bacterial species were, but if on one hand the relationship between \textit{E. coli} and \textit{B. fragilis} has been already studied, the effect of \textit{V. parvula} on \textit{E. coli} has to be investigated yet in the literature. In the UC correlation network, \textit{V. parvula} formed an almost completely isolated module far from the \textit{E. coli}, this result could differentiate the connectome of the UC microbiome from the connectome of the CD microbiome. The isolation of the \textit{E. coli} module in the UC correlation network could represent further the peculiar features of the particular form of IBD. This isolation meant that there are no correlations with the other pathways and the pattern of metagenomic expression across the samples are correlated only inside the same bacterial species. In the NI correlation network, \textit{E. siraeum} and \textit{R. gnavus} pathways were the two main bridge pathways between \textit{E. coli} and the rest of the network, it could be possible to hypothesize that re-establishing a connection between \textit{E. coli} module with the aforementioned bacterial species could lead back to a healthy gut microbiota. In the CD correlation network, \textit{R. intestinalis} pathways had the role of bridge pathways and, in fact, by using the permutation tests between NI and CD samples, we obtained that the most differential pathways were \textit{R. intestinalis} pathways. In the literature, this bacterial species, which has anti-inflammatory properties on the intestinal walls, was depleted in IBD subjects; nevertheless, the complete mechanisms underlying its protective action against IBD are still unknown \cite{Zhu2018RoseburiaColitis, Hoffmann2016MicroorganismsMice}. In the uncommon bacterial pathways of the NI projected network, the \textit{E. coli} module was connected to \textit{the C. boltae}, which, in turn, was linked to \textit{B. longum} module. \textit{B. longum} is a bacterial species that can have anti-inflammatory properties in the human gut \cite{Singh2011CommonDiseases}. By contrast, in the CD projected network, \textit{E. coli} was connected to the rest of the network through two \textit{Dorea longicatena} pathways, which were nodes $680$ and $81$ and were connected to node $697$. On the other hand, in the UC projected network, two \textit{E. coli} nodes were connected to six \textit{C. comes} nodes, showing the existence of an interaction between the two species in the UC diagnosis.

\section*{Methods}

% Topical subheadings are allowed. Authors must ensure that their Methods section includes adequate experimental and characterization data necessary for others in the field to reproduce their work.
\subsection*{Recruitment}
The Inflammatory Bowel Disease Multi'omics Database (IBDMDB) \cite{IBDMDBIBDMDB} is one of the first comprehensive studies of the gut ecosystem's multiple molecular properties involved in the IBD dynamics. Some of the measurements of the microbiome offered by the study are metagenomics, metatranscriptomics and metabolomics.  The data is related to 132 subjects approached in the following five medical centres: Cincinnati Children’s Hospital, Emory University Hospital, Massachusetts General Hospital, Massachusetts General Hospital for Children, and Cedars-Sinai Medical Centre. The patients recruited for the study initially arrived at the hospitals either for routine age-related colorectal cancer screening, presence of other gastrointestinal symptoms, or suspected IBD. The latter could be a consequence of positive imaging, symptoms of chronic diarrhoea or rectal bleeding. If there were no conditions for exclusion right after enrolment, a preliminary colonoscopy was performed to determine study strata. Based on initial analyses, the subjects that were not diagnosed with IBD were labelled as ‘NI’ controls. This group of subjects included the patients who arrived for routine screening and those with more benign or non-specific symptoms \cite{Lloyd-Price2019Multi-omicsDiseases}.

\subsection*{Database}
The IBDMDB website contains the raw, and the final results of the processed information, the complete pipeline for producing the final results is:
\begin{enumerate}
    \item Quality and error checking for completeness, producing \texttt{raw files}.
    \item Anadama pipeline, producing \texttt{products}.
\end{enumerate}
In particular, if we consider the pipeline for producing the metagenomic data, the samples for the quality control process go through the KneadData \cite{KneadDataLab} and the Anadama pipelines. The former is a tool useful to exclude the reads, which are fragments of DNA, related to the host or related to other contaminants from the metagenomic sequencing data, and this separation step is made completely in silico. Whereas the latter, the Anadama pipeline, performs and produces documents from an automated scientific workflow, where a workflow is simply a succession of tasks, such as quantifying operational taxonomic units (OTU). The OTUs are classifications of groups of bacteria closely related to each other by sequence similarity. On the IBDMDB website, there are two versions of data \texttt{Version 2.0} and \texttt{Version 3.0}. \texttt{Version 3.0} has been uploaded with the new version of bioBakery \cite{McIver2018BioBakery:Environment}. In the thesis, we use the \texttt{products} file related to the functional profiles \texttt{Version 2.0}. Moreover, we exploit the HMP2 Metadata file, containing the sample IDs, the subject IDs and the properties associated with each sample. The \texttt{External ID} is the unique ID of the sample, \texttt{Participant ID} is the subject from where the sample has been taken, \texttt{diagnosis} is either ulcerative colitis (UC), Crohn's disease (CD) or control group (NI), \texttt{week\_num} points out the week number, when the sample has been taken and \texttt{data\_type} is the type of sample (metagenomics, 16S, etc.).
we extracted useful information to avoid importing the whole database, and we selected only the samples from the first week (week 0). Moreover, the samples different from metagenomic ones were excluded. Finally, we dropped the samples from the same participant in week 0 and obtained a list of samples ID that were present in both the metagenomic database and the HMP2 Metadata. The metagenomic database contains as row indexes the gene descriptors; specifically, the descriptor is composed of the pathway, genus and species (e.g. "ARO-PWY: chorismate biosynthesis I \textbar g\_\_Alistipes.s\_\_Alistipes\_finegoldii"). To generate the database, the algorithm HUMAnN2 \cite{Franzosa2018Species-levelMetatranscriptomes} has been used. The algorithm can be divided into three phases; firstly, the metagenomic sample is quickly analyzed to seek known species in the gut microbiome. The functional annotation of the identified pangenomes (i.e. the genome of a larger group of species) of the microbiome is concatenated to form a gene database of the sample. Secondly, using this database, the whole sample is aligned, meaning that statistics regarding the species and the genes are made, and unmapped reads are collected. Thirdly, the gene abundances are calculated, and they are combined with the metabolic network to determine the pathways in the microbial community.

\begin{table}[ht]
\centering
\begin{tabular}{|l|l|l|l|} 
\hline
 Pathway & SMPL1 & SMPL2 & SMPL3 \\ 
 \hline
 PWY1 & 0.0435726 & 0.0377424 & 0.0118981 \\ 
 \hline
 PWY2 & 0.0170328 & 0.0144735 & 0.0134886 \\
 \hline
 PWY3 & 0.0145872 & 0.0177172 & 0.0121692 \\
 \hline
 PWY4 & 0.0545121 & 0.0018744 & 0.0175601 \\
 \hline
 PWY5 & 0.0881223 & 0.0111788 & 0.0163441 \\
 \hline
\end{tabular}
\caption{Representation of the database containing the gene expression data with example values.\label{table:1}}
\end{table}

To reduce the number of genes present in the resulting network, we built the correlation matrices for three different groups; namely, one group for the pathways present in a percentage between 25\% and 50\% of the subjects (uncommon pathways), another group for those in the range between 50\% and 75\% (common pathways), and lastly a group for the pathways present in more than 75\% (prevalent pathways). 
\subsection*{Correlation Networks}
Correlation networks are built from the following steps:
\begin{enumerate}
    \item pairwise gene similarity score (correlation);
    \item thresholding.
\end{enumerate}
Normalization methods, correlation measures (Pearson or Spearman), significance and relevance are still debated \cite{Tieri2019NetworkBioinformatics}. In our work, we chose Pearson correlation similarly to \cite{MacMahon2013CommunityMatrices}. 

To transform a correlation matrix into the correlation network, we used a thresholding method inspired by a brain network technique that was used to cut the least important edges and keep the significant relationships among the nodes, hence, we calculated the absolute value of the correlations, making the signs irrelevant. This method consists of increasing a cut-off threshold until the network connectivity breaks apart; because of this property, this cut-off threshold is also known as the percolation threshold. This method has been considered one of the most effective methods to maximise the information quantity kept by the network \cite{Nicolini2020Scale-resolvedEntropy}.  In our work, we started from a cut-off threshold of $t=0$, and we used a bisection method to get to the percolation threshold. In the bisection method, we flattened the absolute values in the weighted adjacency matrix into a sorted array, we chose the median value and used it as the new cut-off threshold, we calculated the connectivity of the graph built from the adjacency matrix having this cut-off threshold, finally, if we obtained a connected graph with the median value as a cut-off threshold, we used as the sorted array the upper half array, on the contrary, we used the lower half. The procedure was iterative until convergence which corresponded to an array with zero length or with the same head and same tail.

\subsection*{Co-expression Network from Bipartite Projected Networks}
Bipartite networks are graphs $G=(U, V, E)$ where the nodes can be divided into two disjoint sets, $U$ and $V$, and every edge in $E$ links an element in $U$ to an element in $V$. In our work, we designated with $U$ the set of nodes representing the genes and with $V$ the set of nodes representing the samples. We can find in $E$ all the edges connecting the gene $u$ to the sample $v$ if the gene was over-expressed in the corresponding sample. We evaluated the over-expression of a gene through the binarization of the data that led to the construction of a biadjacency matrix $B$ of size $|U| \times |V|$ that described the bipartite network $G=(U, V, E)$ with the entries $(0,1)$, where $B_{ij}=1$ if the gene $v_i$ is over-expressed in the sample $u_j$, and $B_{ij}=0$ if it is not over-expressed. Biadjacency matrices are rectangular matrices where on one side there are the nodes in $U$, and on the other side the nodes in $V$.
\subsubsection* {Binarization}
The binarization process is useful to highlight the edges in $E$ that are over-expressed. By using the revealed comparative advantage (RCA) \cite{Balassa1965Trade1}, We highlighted the over-expressed genes for specific samples:

\begin{equation}
    RCA_{ij}=\frac{E_{ij}/\sum_{j'\in V} E_{ij'}}{\sum_{i'\in U} E_{i'j}/\sum_{i'\in U, j'\in V} E_{i'j'}}
\end{equation}
where $E$ is the expression of the gene $i$ in the sample $j$. When $RCA_{ij}>1$, the quantity of gene $i$ in sample $j$ can be considered over-expressed and the entry $b_{ij}=1$, in the other case $RCA_{ij} \leq 1$, then $b_{ij}=0$.

\subsubsection*{Randomization of bipartite networks}
To generate a null model useful to calculate the statistically important properties of a real bipartite network, we randomized the bipartite networks by using the package BiCM \cite{Bruno2020BiCM:Model.}. In particular, the package is based on the works by \cite{Squartini2011AnalyticalNetworks}, \cite{Saracco2015RandomizingWeb}, and \cite{Saracco2017InferringApproach}. In the aforementioned works, the Shannon entropy defined as
\begin{equation}
    \mathcal{S}=-\sum_{\mathbf{M}\in\mathcal{G}}P(\mathbf{M})\ln{P(\mathbf{M})} 
\end{equation}
is maximized, where $\mathcal{G}$ is an ensemble of binary, undirected, bipartite networks, and $\overrightarrow{C}(M)$ is a given set of constraints. The result is:
\begin{equation}
    P\left(\mathbf{M}|\overrightarrow{\theta}\right)=\frac{e^{-H\left(\mathbf{M},\overrightarrow{\theta}\right)}}{Z\left(\overrightarrow{\theta}\right)}
\end{equation}

Where $H\left(\mathbf{M},\overrightarrow{\theta}\right)=\overrightarrow{\theta}\cdot \overrightarrow{C}(M)$ is the hamiltonian and $Z\left(\overrightarrow{\theta}\right)=\sum_{M\in G}e^{-H\left(\mathbf{M},\overrightarrow{\theta}\right)}$ is the normalization. In the case of the bipartite extension of the configuration model (BiCM), the hamiltonian becomes:
\begin{equation}
    H\left(\mathbf{M},\overrightarrow{\theta}\right)=\overrightarrow{\alpha}\cdot\overrightarrow{d}(\mathbf{M})+\overrightarrow{\beta}\overrightarrow{u}(\mathbf{M})
\end{equation}
because we have two layers of nodes and we constrained the degree sequences $\overrightarrow{d}(\mathbf{M})$ and $\overrightarrow{u}(\mathbf{M})$. $\overrightarrow{d}(\mathbf{M})$ is the degree sequence of the genes and $\overrightarrow{u}(\mathbf{M})$ is the degree sequence of the samples.
\subsubsection*{Projection}
One way to compress the information contained in a bipartite network is to project the bipartite network onto one of the two layers (gene/pathway layer or sample layer). We carried out the projection by connecting in the same layer the nodes that were linked by a common node in the other layer. The projection leads to a loss of information itself, so to avoid further loss of information, we weighted the edges by the number of common nodes neighbouring the nodes in the same layer \cite{Neal2014TheCo-behaviors}. The algorithm to perform the projection is:
\begin{enumerate}
    \item select the partition on which the projection will be done
    \item take two nodes of the selected partition, $n$ and $n'$, and calculate their similarity
    \item by evaluating the corresponding $p$-value compute the statistical significance of the calculated similarity with respect to a properly-defined null model;
    \item if, and only if, the $p$-value associated with the link $n$ and $n'$ is statistically significant, connect the selected nodes.
\end{enumerate}
The similarity in the second step of the algorithm is evaluated by:
\begin{equation}
    V_{nn'}=\sum_{c=1}^{N_c}m_{nc}m_{n'c}=\sum_{c=1}^{N_c}V_{nn'}^c,
\end{equation}
where $V_{nn'}^c \equiv m_{nc}m_{n'c}$ and it is clear from the definition that $V_{nn'}^c = 1$ if, and only if, both $n$ and $n'$ are common neighbours of $c$. The third step of the algorithm passes through the calculation of the $p$-value of the Poisson–Binomial distribution, i.e. the probability of observing a number of V-motifs greater than, or equal to, the observed one (which will be indicated as $V_{nn'}^*$:
\begin{equation}
    p-value(V_{nn'}^*)=\sum_{V_{nn'}\ge V_{nn'}^*}f_{PB}(V_{nn'}) = 1 - \sum_{V_{nn'}\le V_{nn'}^*}f_{PB}(V_{nn'}).
\end{equation}
Finally, in the last step of the algorithm, in order to understand which $p$-values were significant, a false-discovery rate or FDR  has been adopted to take into account the fact that we were testing multiple hypotheses \cite{Benjamini1995ControllingTesting}.

\subsection*{Betweenness centrality}
There are many different centrality measures in network science; these measures describe the importance of a node in the network. The betweenness centrality was introduced by Freeman \cite{Freeman1977ABetweenness}, and it considers more important the nodes that behave as bridges in the network. It can be calculated as:
\begin{equation}
    C_B(i)=\sum_{s \ne t \ne i \in V} \frac{\sigma_{st}(i)}{\sigma_{st}} 
\end{equation}
where $\sigma_{st}$ is the number of shortest paths connecting $s$ and $t$, whilst $\sigma_{st}(i)$ is the number of shortest paths connecting $s$ and $t$ and going through $i$. %\cite{Brandes2001ACentrality}.

\subsection*{Community detection}
In the study of network science, both natural complex networks and artificial complex networks display a modular behaviour, i.e. groups of nodes are more densely connected within the members of the group than with the rest of the network. This phenomenon can also be described by a function called modularity \cite{Newman2006ModularityNetworks}, which can be used as a parameter for one of the several ways to perform community detection in complex networks. In our work, we used Louvain method \cite{Blondel2008FastNetworks} because it is suited to large complex networks. Louvain method is based on an optimization problem that can be solved in a time $O(n \cdot log_2 n)$ where $n$ is the number of nodes in the network \cite{Lancichinetti2009CommunityAnalysis}. The method is based on the aforementioned modularity optimization. The modularity is defined as \cite{Newman2004AnalysisNetworks},
\begin{equation}
    Q=\frac{1}{2m}\sum_{i,j}\left[A_{ij}- \frac{k_i k_j}{2m} \right]\delta(c_i,c_j).
\end{equation}
The algorithm is based on two phases that repeat iteratively. In the first phase, each node is repeatedly moved individually between the communities to maximize modularity. The first phase stops when no further individual move can improve the modularity. In the second phase, each community formed in the first phase is considered as a node of a weighted graph, where the weights of the edges are given by the sum of the edges connecting the nodes in the communities. The algorithm has a high efficiency partly because the gain modularity $\Delta Q$, due to moving a node $i$ into a community $C$, can be steadily calculated as:
\begin{equation}
    \Delta Q = \left[\frac{\sum_{in}+k_{i,in}}{2m}-\left(\frac{\sum_{tot}+k_{i}}{2m}\right)^2\right]-\left[\frac{\sum_{in}}{2m} -\left(\frac{\sum_{tot}}{2m}\right)^2 -\left(\frac{k_{i}}{2m}\right)^2 \right],
\end{equation}
where $\sum_{in}$ is the sum of the weights of the edges inside $C$, $\sum_{tot}$ is the sum of the weights of the edges going from the outside to the nodes inside $C$, $k_{i}$ is the sum of the weights of the edges going to node $i$, and $k_{i,in}$ is the sum of the weights of the edges going from $i$ to the nodes in $C$ and, finally, $m$ is the sum of the weights of all the edges in the graph. One of the limitations of community detection based on the modularity is the resolution limit \cite{Fortunato2007ResolutionDetection}. This limitation to modularity may be present when $l_s \approx \sqrt{2L}$, where $l_s$ is the number of internal links in a module $S$ and $L$ is the total number of links in the network and it can be overcome through several methods, one of the most promising is Surprise maximization \cite{Aldecoa2013SurpriseNetworks}.

\bibliography{sample}

\noindent % LaTeX formats citations and references automatically using the bibliography records in your .bib file, which you can edit via the project menu. Use the cite command for an inline citation, e.g.  \cite{Hao:gidmaps:2014}.

% For data citations of datasets uploaded to e.g. \emph{figshare}, please use the \verb|howpublished| option in the bib entry to specify the platform and the link, as in the \verb|Hao:gidmaps:2014| example in the sample bibliography file.

%\section*{Acknowledgements (not compulsory)}

%Acknowledgements should be brief, and should not include thanks to anonymous referees and editors, or effusive comments. Grant or contribution numbers may be acknowledged.

\section*{Author contributions statement}

Conceptualization, G.C., T.G. and M.H.; methodology, G.C., T.G. and M.H.; software, M.H. and T.G.; validation, M.H. and T.G.; formal analysis, M.H.; investigation, M.H.; resources, M.H.; data curation, M.H.; writing---original draft preparation, M.H.; writing---review and editing, T.G. and G.C.; visualization, M.H.; supervision, M.H., T.G. and G.C.; project administration, G.C.; funding acquisition, G.C. All authors have read and agreed to the published version of the manuscript.

\section*{Additional information}

To include, in this order: \textbf{Competing interests} The authors declare no conflict of interest. 

The corresponding author is responsible for submitting a \href{http://www.nature.com/srep/policies/index.html#competing}{competing interests statement} on behalf of all authors of the paper. This statement must be included in the submitted article file.

%\begin{figure}[ht]
%\centering
%\includegraphics[width=\linewidth]{stream}
%\caption{Legend (350 words max). Example legend text.}
%\label{fig:stream}
%\end{figure}

%\begin{table}[ht]
%\centering
%\begin{tabular}{|l|l|l|}
%\hline
%Condition & n & p \\
%\hline
%A & 5 & 0.1 \\
%\hline
%B & 10 & 0.01 \\
%\hline
%\end{tabular}
%\caption{\label{tab:example}Legend (350 words max). Example legend text.}
%\end{table}

%Figures and tables can be referenced in LaTeX using the ref command, e.g. Figure \ref{fig:stream} and Table \ref{tab:example}.

\end{document}